\documentclass[twocolumn,showpacs,preprintnumbers,amsmath,amssymb]{revtex4}

\usepackage{graphicx}
\usepackage{dcolumn}
\usepackage{bm}

\newcommand\Tr{\mbox{Tr}}            

\begin{document}


\title{Small scales and anisotropy in low \textit{Rm} MHD turbulence}

\author{A. Poth\'erat}
\email{ap312@eng.cam.ac.uk}
\author{T. Alboussi\`ere}%
\email{ta209@eng.cam.ac.uk}
\affiliation{%
 Cambridge University Engineering Department\\
 Trumpington street, Cambridge CB2 1PZ
}%


\date{24 June, 2003}		










\begin{abstract}
In this paper, we derive estimates for size of the small scales and the 
attractor dimension in low $Rm$ magnetohydrodynamic turbulence by deriving 
a rigorous upper bound of the dimension of the attractor representing this flow.
To this end, we find an upper bound for the maximum growth rate of any $n$-dimensional 
volume of the phase space by the evolution operator associated to the 
Navier-Stokes equations. As explained in \cite{constantin85_jfm}, The value of 
$n$ for which this maximum is zero is an upper bound for the attractor 
dimension. In order to use this property in the more precise
case of a 3D periodical domain, we are led to calculate the distribution of 
$n$ modes which minimises the total (viscous and Joule) dissipation. This set of
modes turns out to exhibit most of the well known properties of MHD turbulence, 
previously obtained by heuristic considerations such as the existence of the 
Joule cone under strong magnetic field. The sought estimates for the small 
scales and attractor dimension are then obtained under no physical assumption 
as functions of the Hartmann and the Reynolds numbers and match the Hartmann
number dependency of heuristic results. A necessary condition for the flow to 
be tridimensional and anisotropic (as opposed to purely two-dimensional) is also built.

\end{abstract}

\pacs{Valid PACS appear here}

\maketitle

\section{Introduction}
\label{sec:intro}
MHD Turbulence at low magnetic Reynolds number \textit{Rm} (\textit{i.e.} 
for which the magnetic field is
not disturbed by the flow) is of great interest for laboratory experiments as 
well as for industrial applications including metallurgy and the study of liquid metal 
blankets used in nuclear fusion reactors. It essentially differs from classical
hydrodynamic turbulence by  the additional Joule dissipation arising from
 the electric currents present in the flow \cite{davidson01,buhler01}. This anisotropic dissipation 
 competes with the usual viscous dissipation and when it is dominant, the flow 
 exhibits very  characteristic features: first, the 
 turbulent modes are confined outside  the so-called Joule cone in the Fourier
 space (of axis the  direction of the applied magnetic field, and the angle of which is
 governed by the ratio of the Lorentz to the inertial forces). Also the 
 additional dissipation leads to a faster energy decay proportional to 
 $t^{-1/2}$ for freely decaying turbulence \cite{moffatt67} when it is much 
 greater than viscous dissipation. Homogeneous 3D MHD
 turbulence also 
 exhibits a $k^{-3}$ power density spectrum, different from the usual $k^{-5/3}$
  law. This spectrum has been observed experimentally and 
  heuristic considerations suggest it result from a local balance between inertia and 
  Lorentz forces \cite{alemany79}. One of the most 
  striking features of low-$Rm$ MHD turbulence is its anisotropy due to the fact that 
  vortices stretched along the magnetic field lines escape ohmic dissipation \cite{sm82, 
  dav97}, and which results in the existence of the Joule cone in Fourier 
  space where the modes are strongly dissipated. The flow may then become two-
  dimensional  when the modes with a non-zero wave number component in the 
  magnetic field direction are all killed by Joule dissipation.\\
  To summarise, three points characterise well low $Rm$-MHD turbulence (this does not 
  extend to moderate and high   $Rm$ MHD turbulence): its faster decay, 
  its anisotropy, and its power density spectrum. Although 
  all those quantities are known through experiments and relate well to 
  heuristic considerations, none of them is clearly linked to the 
  mathematical
  properties of the Navier-Stokes equations. This point is important for two 
  reasons: first, results derived from mathematical properties of the equations
 are very robust and therefore give 
 indisputable support to heuristic arguments, should they match. Secondly, 
 they provide some deep insights into the behaviour of the solutions, which is 
 necessary when one wants to undertake  calculations on turbulent flows. 
 We aim at doing a step towards filling this
 gap by studying anisotropy and small scales in a fully established 
 turbulent flow, by means of the theory of dynamical systems. This latter tool
 is indeed very suitable to understand turbulence as some of its objects are 
 in direct relation with characteristic properties of turbulence such as the
 size of the smallest scales which is  expressed by  the idea that the solutions of Navier-
Stokes are described by a finite (but possibly large) number of determining 
  modes. This number is also of the same order of magnitude as the dimension
  of the attractor of the system for which estimates can be found.
 Some important questions then arise: 1) how many
  modes are required to describe the flow? 2)which modes? and 3) what 
  information is lost if one attempts a calculation using a smaller number of
 modes? The purpose of this paper is to suggest some ideas 
 for  1) and 2) in the case of MHD turbulence. A way to answer 
 1) is to find an upper bound for the
  dimension of the attractor of the dynamical system formed by the anisotropic 
  Navier-Stokes
   equation (\textit{i.e} with Lorentz force) and associated boundary conditions. 
   This work has already been  carried 
  out without magnetic field by \cite{constantin88} who found a close 
  bound for the attractor dimension of the 2D problem under the form 
  $\mathcal G ^{2/3}(1+\ln \mathcal G)^{1/3}$ ($\mathcal G$ is the Grashof 
  number based on a measure of the applied forcing), which fits well with the typical
  size of the small scales given by \cite{kraichnan80} from heuristic 
  considerations. A similar result has been found in 3D by \cite{constantin85_ams} and
  summarised in \cite{constantin85_jfm}
  but  the final $Re^3$ bound found for the attractor dimension ($Re$ is the Reynolds 
  number) is not as  sharp as the previous bound, when compared to the
  $Re^{9/4}$ estimate by the Kolmogorov theory. However, some estimates for the
  inertial terms derived from this reference will be used to tackle the MHD 
  problem. Note that a thorough study of the general MHD equations 
  (\textit{i.e.} the system formed with the Navier-Stokes equations and the 
  induction equations, which covers situations where  velocity and magnetic 
  field are fully coupled) is presented in \cite{sermange83}. In particular, 
  it is shown
  that any invariant set for this system (hence any attractor) has a finite 
  Hausdorf dimension. Note also that MHD turbulence where the magnetic field
  can fluctuate has been widely studied and even if the physical mechanisms
  involved are very different to those in the 
  case we study here, a similar anisotropy is observed when a mean magnetic 
  field is imposed (see for instance \cite{shebalin83})\\
  The layout of the paper is as follows: we first review the tools of system 
  dynamics used thereafter (\textit{i.e.} the method for calculating the 
  attractor dimension) and show how they relate to our problem. It turns out 
  that calculating the upper bound for the attractor dimension is related to the
  problem of finding the least dissipative modes.
  Section 2 is devoted to finding those modes and their properties as well
  as the upper bound itself: those modes are
  found to correspond to prominent modes of actual MHD-turbulent flows. In 
  section 3,
  analytical estimates are given and comparison is drawn with the usual 
  heuristic arguments.
\section{Navier-Stokes equations and Dynamical systems}
\label{sec:systdyn}
\subsection{Method for calculating an upper bound for the attractor dimension of a dynamical system}
\label{sec:method}
We shall now give some guidelines about the method which we use to derive  such an upper bound. A 
dynamical system with vector valuated unknown $\mathbf{x}$ is defined by an evolution equation of the 
form:
\begin{equation}
\frac{d \mathbf x}{dt}=F(\mathbf x)
\label{eq:syst}
\end{equation}
together with boundary conditions on the considered domain spanned by the variable $\mathbf x$ 
or phase space. By definition of a global attractor for the system (which is a set located in the 
phase space), a solution of (\ref{eq:syst}) always ends up being arbitrary close to it at infinite 
time.  Therefore, if we consider any set of $n$ infinitesimal independent departures from a solution located in the 
attractor $(\delta \mathbf x_k)_{k=1...n}$, the subset of the phase space generated by these disturbances 
will eventually end up 
within the attractor in the limit $t \rightarrow \infty$. This implies that if the initial dimension $n$ 
of this subset is greater than the attractor dimension, its $n$-dimensional volume tends to zero at 
infinite time (as, for instance, a 3D cube would have to become "flat", \textit{i.e.} of volume $0$,  
in order to fit in a plane at infinite time). Therefore, the lowest value of $n$ for which the volume of 
the subspace generated by any set of $n$ disturbances  annihilates at infinite time is an upper bound for the
attractor dimension. This result is expressed rigorously and extended to non-integer values of $n$ by the 
theorem of Constantin and Foias (\cite{constantin87}).\\
In order to be able to use this theorem, it suffices to find the lowest value of $n$ which corresponds 
to a zero value of the maximum expansion rate among all possible $n$-dimensional infinitesimal disturbances.  The evolution
of each disturbance is 
expressed by  linearisation of (\ref{eq:syst}) in the vicinity of the attractor:
\begin{equation}
\frac{d}{dt} \delta \mathbf x =\mathcal A \delta \mathbf x+O(\delta \mathbf x^2)
\label{eq:syst_var}
\end{equation}
Then the expansion rate of the $n$-volume 
$V_n=\|\delta\mathbf x_1 \times  ...\times \delta\mathbf x_n\|$ 
is the sum of the expansion rates in all the eigendirections of $\mathcal A$ within the $n$-dimensional
 subspace:
\begin{equation}
V_n(t)
=V_n(t=0)\exp(t\langle\Tr [\mathcal{A}\mathbf{P}_n]\rangle_t)
\end{equation}
\begin{figure}
\centering
\includegraphics[scale=0.4]{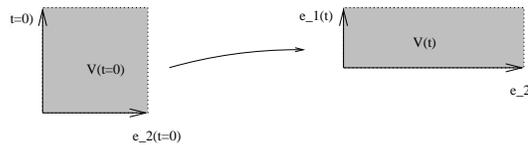}
\caption{Evolution of the volume associated with a base of orthogonal eigenvectors of an operator
in the phase space. The dimension $n$ is set to 2.}
\label{fig:evol}
\end{figure}
where $\mathbf P_n$ stands for the projector onto the $n$ dimensional subspace spanned by 
$(\delta \mathbf x_k(t))_{k=1...n}$, and $\langle \rangle_t$ stands for the longest possible time-average. 
One can get an idea of how this result comes up by considering the volume spanned by a base 
of orthogonal eigenvectors of $\mathcal A$ 
$(\mathbf e_k)_{k \in \{1..n\}}$, in the case where $\mathcal A$ is self-adjoint and 
time-independent. If 
$(\lambda_k)_{k \in \{1..n\}}$ is the related set of eigenvalues, then the length of the volume 
element along the direction $\mathbf e_k$ evolves as (see figure \ref{fig:evol}):
\begin{equation}
\mathbf e_k(t)=\mathbf e_k(t=0)\exp(\lambda_k t)
\end{equation}
As the $(\mathbf e_k)$ are orthogonal, the volume is simply the product of the lengths in all directions, 
so that it evolves as:
\begin{eqnarray}
V_n&=&\|\mathbf e_1(t=0)\|\|\mathbf e_2(t=0)\|...\|\mathbf e_n(t=0)\| \exp(\sum_{k=1..n} \lambda_k t)
\nonumber \\
&=&V_n(t=0)\exp(\Tr\mathcal A t)
\end{eqnarray}

Finding the maximum expansion rate over every possible $n$-dimensional subspace then comes down to finding 
the maximum of the trace of the linearised evolution operator over all possible sets of $n$  of its 
eigenmodes. Let us now apply these ideas to the problem of MHD turbulence.

\subsection{The Navier-Stokes equations as a dynamical system}
Let us consider an incompressible electrically conducting fluid in a finite domain, subject to
a permanent, uniform magnetic field $\mathbf B$ aligned with the $z$-axis . 
If $\sigma$ is the electrical 
conductivity, $\rho$ is the density, $\nu$ is the kinematic viscosity, the motion equations for 
velocity $\mathbf u$, pressure $p$ electric current density $\mathbf j$ can be written:
\begin{eqnarray}
(\partial_t + \mathbf u .\nabla)\mathbf u + \frac{1}{\rho}\nabla p
&=& \nu (\nabla^2 \mathbf  u + \frac{1}{\rho \nu} \mathbf j \times \mathbf B) + \mathbf f
\label{eq:ns}\\
\nabla . \mathbf u &=& 0
\end{eqnarray}
where $\mathbf f$ represents some forcing independent of the velocity field.
The set of Maxwell equations as well as electric current conservation and the Ohm's law 
are normally required to close the system. However, we assume here that the magnetic field is
not disturbed by the flow. In other words, the magnetic diffusion is supposed to take place
instantaneously at the time scale of the flow ("low magnetic Reynolds number" approximation).
In this case, \cite{roberts67} has shown that the Lorentz force decomposes
as the sum of a magnetic pressure term and a rotational term:
\begin{equation}
\mathbf j \times \mathbf B= 
\frac{\nu}{\rho} \nabla p_m + \frac{\sigma B^2}{\rho \nu} \nabla^{-2} \partial^2_{zz}\mathbf u.
\label{eq:roberts}
\end{equation}
This reveals the nature of the electromagnetic effects on the flow: the first term accounts 
for the electromagnetic pressure (of little effect in incompressible flows). The second 
term can be interpreted as a momentum diffusion along the magnetic field lines (\cite{sm82}) 
which tends to homogenise $x,y$ components of the  velocity  along $z$. This 
stretches vortices along the $z$ direction. The actual turbulent flow therefore exhibits 
some anisotropy which results from the competition between this momentum diffusion and the 
tendency from inertial terms to favour return to isotropy. Note that if the electromagnetic 
effects are
dominant, the stretched vortices can reach the boundaries of the flow, which then becomes 
two-dimensional.\\
Injecting (\ref{eq:roberts}) in the Navier-Stokes equation (\ref{eq:ns}), the electromagnetic pressure 
is absorbed in the hydrodynamic pressure term so that the entire MHD problem is expressed using the velocity
only. The related variation equation which governs the evolution of a three dimensional 
perturbation $\delta \mathbf u$ of the solution $\mathbf u$ then takes the form:
\begin{eqnarray}
\partial_t \delta \mathbf u
&=& \underbrace{-\mathbf u .\nabla \delta \mathbf u - \delta \mathbf u .\nabla\mathbf  u}_{non-linear inertia}
\nonumber \\
+ \underbrace{\nu (\mathbf \nabla^{-2}  
+\frac{\sigma B^2}{\rho \nu} \nabla^{-2} \partial^2_{zz})\delta \mathbf u }_{dissipation}
\label{eq:ns_var}\\
\nabla . \delta \mathbf u &=& 0
\end{eqnarray}
In the literature, the non- linear inertial terms are often written as a bilinear operator
$\mathcal B(\mathbf u,\delta \mathbf u)$, and the dissipation, as a linear
 operator that we call $\mathcal D_{Ha}$ (As it will be seen to depend on the 
 Hartmann number $Ha$ in section \ref{sec:dissipation}).
One can guess from this equation, that the evolution of small volume of the phase space 
generated by a set of $n$  disturbances
(as defined in section \ref{sec:method}) results from the competition between inertial terms which 
tend to expand the volume by vortex stretching and dissipative terms
which tends to damp the disturbances, and hence reduce the volume.\\
The case without magnetic field has been investigated in 2 and 3 dimensions. In 2d, 
\cite{doering95} found an upper bound for the attractor dimension which matches well 
the results obtained by Kolmogorov-like arguments:
\begin{equation}
d_{2d} \leq c_1 \mathcal G^{2/3}(1+ln \mathcal G)^{1/3}
\label{eq:dim2d}
\end{equation}
where $\mathcal G$ is the Grashof number expressing the ratio of the forcing to the viscous friction
and $c_1$, as well as  every $c_i$ introduced throughout the rest of the paper, are constants of order 1.
To this day, no rigorous estimate for the attractor dimension of the 3D problem precisely matches Kolmogorov's
prediction for the number of degrees of freedom.
One of the main reasons is that unlike in 2D, it has not
yet been proved that the velocity gradients remain finite at finite time, which lets the door open
to possible singularities. However, one can work under the assumption that the flow remains regular
at finite time and define the maximum local energy dissipation rate as:
\begin{equation}
\epsilon = \nu \langle \sup_{\mathbf u} \sup_{\mathbf r} \|\nabla \mathbf u (\mathbf r,t)\|^2 \rangle_t
\label{eq:eps}
\end{equation}
One can also define a 
Reynolds number $Re$ using a suitable velocity scale and a typical large scale $L$,
which can be extracted from the eigenvalue 
of the laplacian of smallest module $\lambda_1$, such that $L=\lambda_1^{-1/2}$:
\begin{equation}
Re=\frac{ L \langle \sup_{\mathbf u} \sup_{\mathbf r} \| \mathbf u (\mathbf x,t)\|^2 \rangle^{1/2}_t}
{\nu}
\end{equation}
Here, $\sup_{\mathbf u}$ stands for the upper bound over the set of solutions $\mathbf u$ in the phase space,
whereas $\sup_{\mathbf r}$ stands for the upper bound over the physical domain. Note that fixing the value of the 
Reynolds number is the 3D equivalent to fixing the value of the Grashof number, which represents the forcing in 2D.
It should be underlined at this point, that as the attractor is only defined for quasi-steady states, it is entirely
determined by the balance between forcing (given by the value of the Reynolds number) and dissipation (the nature of
which is fixed by the value of the Hartmann number).\\
Under this assumption that the velocity remains finite,
an upper bound for the trace of the 
operator $\mathcal B(.,\mathbf u)$ on any $n$-dimensional subspace of the phase space is presented in \cite{constantin85_jfm}:
\begin{equation}
|\Tr(\mathcal B(.,\mathbf u))|< \frac{1}{2}\nu \lambda_1 n Re^2
\label{eq:trb}
\end{equation}
Also, studying the sequence of eigenvalues of the dissipation operator (which reduces to a Laplacian
in the absence of magnetic field) on a finite physical domain with appropriate boundary conditions,
gives access to the trace of the dissipation operator (see for instance \cite{doering95})
and provides an upper bound for the trace of the total evolution operator, on any $n$-dimensional
subspace of the phase space:
\begin{equation}
\Tr((\mathcal B(.,\mathbf u)+ \nu \nabla^2)\mathbf P_n) \leq \nu \lambda_1 n (\frac{1}{2}Re^2-c_2 n^\frac{2}{3})
\label{eq:trace_constantin}
\end{equation}
One can be sure that when $n$ is such that the \textit{r.h.s.} of (\ref{eq:trace_constantin}) is 
negative, all $n$-volumes shrink, hence $n>d_{3D}$ where $d_{3D}$ is the attractor's dimension 
(this is Constantin and Foias theorem \cite{constantin85_cpam}. It then comes from  (\ref{eq:trace_constantin}) that:
\begin{equation}
d_{3d} \leq c_3 Re^3
\label{eq:dim3d}
\end{equation}
The bound (\ref{eq:dim3d}) is a rather loose estimate when compared to the $Re^{9/4}$ number of degrees 
of freedom
derived from Kolmogorov arguments which assumes the existence of a power-law spectrum and uses a Reynolds 
number defined on  the mean-square velocity. This is probably due to the difficulty in getting
estimates for the norms of the velocity gradients, as well as to the fact that the bound given here
does not rely on the existence of a power-law spectrum, which makes it also valid for low values of $Re$, 
unlike the K41 \cite{k41} theory.\\
Coming back to the problem of finding an upper bound for the  attractor of an MHD turbulent flow under 
imposed magnetic field, our task now consists mainly in finding  an upper bound for the trace of the 
operator $\mathcal D_{Ha}$ on any $n$ dimensional subspace, as the estimate for the inertial terms 
(\ref{eq:trb}) can then still be used to derive the minimum of the trace of the linearised evolution operator.
The study of the dissipation operator, with the aim of finding such a minimum is the purpose of section
\ref{sec:dissipation}. To this end, and in order to keep the calculations simple,  we shall restrict 
the problem to a physical domain defined by a three-dimensional periodic box of size $2\pi L$.
\section{Modes minimising the dissipation}
\label{sec:dissipation}
\subsection{Eigenvalue problem for the dissipation operator}
\label{sec:eigenval}
{We now look for the maximum trace of the dissipation operator, or bearing in mind that this trace is 
negative, we aim at finding the modes with the least dissipation. The physical domain is a 3d-periodic 
box of size $2\pi L$ in a uniform, vertical magnetic field.
Normalising distances by $L$, the dissipation operator rewrites $\mathcal D_{Ha}=
\frac{\nu}{L^2}(\nabla^2+Ha^2\nabla^{-2}\partial^2_{zz})$, where the square of the Hartmann number 
$Ha=LB\sqrt \frac{\sigma}{\rho \nu}$ represents the ratio of Joule to viscous dissipation at the 
largest scale $L$. From now on,  $\mathcal D_{Ha}$ will denote the non-dimensional form of the dissipation
operator, normalised by $\frac{\nu}{L^2}$ It is straightforward to see that under periodic boundary conditions, 
the laplacian is invertible so that $\mathcal D_{Ha}$ is also invertible, as well as compact and self-adjoint.
 $\mathcal D_{Ha}$ therefore has a discrete spectrum. Finding the minimum value of the 
modulus of the trace of  $\mathcal D_{Ha}$ over any $n$-subspace then comes down to finding the $n$
eigenvalues of $\mathcal D_{Ha}$ of smallest module $(\lambda_k)_{k=1..n}$. In other words, 
we need to find the $n$
least dissipative modes. The rest of this subsection is devoted to this task.\\
The eigenvalues problem for $\mathcal D_{Ha}$ can be written:
\begin{eqnarray}
(\nabla^4+Ha^2 \partial^2_{zz})\mathbf v&=& \lambda\nabla^2 \mathbf v
\label{eq:eigenprob}\\
\nabla . \mathbf v&=&0
\end{eqnarray}
Under periodic boundary conditions, $\frac{\partial}{\partial x}$, $\frac{\partial}{\partial y}$ and
$\frac{\partial}{\partial z}$ commute with $\mathcal D_{Ha}$ so that each component $v_x$, $v_y$ and $v_z$ of
the solution $\mathbf v=(v_i)_{i\in\{x,y,z\}}$ of (\ref{eq:eigenprob}) is of the form:
%
%
%
\begin{equation}
v_i(\mathbf x)=V_i\exp(\mathbf {k. x}+\phi_i)
\label{eq:form_vi}
\end{equation}
with
\begin{eqnarray}
\mathbf k&=& (k_x,k_y,k_z)\in \mathbb Z^3,
\nonumber \\ 
\mathbf x&=&(x,y,z)\in [0..2\pi L]^3,
\nonumber\\
\phi_i &\in& [-\pi, \pi]
\nonumber
\end{eqnarray}
Note that $\mathbf k \neq 0$, as $\mathcal D_{Ha}$ is invertible.
The continuity equation implies that $\mathbf{k.V}=0$ so that eventually, the dimension of the 
eigenspace associated to $\mathbf k$ is 2.
Wavenumbers $k_x,k_y,k_z$ are related to the eigenvalue $\lambda$ through the dissipation equation, obtained by
injecting (\ref{eq:form_vi}) in (\ref{eq:eigenprob}):
\begin{equation}
\lambda(k_x,k_y,k_z)=-(k_x^2+k_y^2+k_z^2)-Ha^2\frac{k_z^2}{k_x^2+k_y^2+k_z^2}
\label{eq:dissip}
\end{equation}
We shall now assume that the components of $\mathbf k$ are positive and that one $\mathbf k$ 
actually represents 8 (\textit{resp.} 4 \textit{resp.} 2) different modes if 
$\mathbf k$ has no (\textit{resp.} one \textit{resp.} two) zero component(s), 
 so that the eigenspace associated to $\lambda(k_x,k_y,k_z)$ 
has a dimension 16 (\textit{resp.} 8 \textit{resp.} 4).
Each eigenvalue $\lambda(k_x,k_y,k_z)$ can be interpreted as the dissipation rate associated 
with the mode $(k_x,k_y,k_z)$. We see that because of Joule dissipation, the total dissipation 
is always higher than the viscous dissipation alone (obtained for $Ha=0$). Note also that as the 
eigenmodes are trigonometric functions, the space spanned by $(k_x, k_y,k_z)$ is the discrete 
Fourier space.\\
\subsection{Distribution of the least dissipative modes in the Fourier space}
\label{sec:distrib}
The $n$ least dissipative modes are given by the $n$ lowest values of 
$-\lambda(k_x,k_y,k_z)$. In order to find them, we note that $k_z\mapsto-\lambda(k_x,k_y,k_z)$ 
is always increasing. This implies that the $n$ minimal modes have to be located "below" 
(\textit{i.e.} closer to the $(k_x,k_y)$ plane than... ) the manifold $\lambda(k_x,k_y,k_z)=\lambda_m$
 where $-\lambda_m(n)$ is the maximum value of $-\lambda(k_x,k_y,k_z)$ reached on the set of $n$ minimal modes.
The fact that $\partial-\lambda/\partial k_z >0$ also implies that all $\mathbf k$ inside the 
volume defined by this curve, $k_x \geq 0$, $k_y \geq 0$
 and $k_z \geq 0$, do belong to the set of $n$  minimal modes 
 \footnote{In the case where several triplets achieve the value $\lambda_m$, the curve
$f=\lambda-\lambda_m$, may in fact enclose more than $n$ points, but this little error is of
no consequence for our purpose, and is anyway addressed in the numerical method described in section
\ref{sec:numeric}}. 
\begin{figure}
\centering
\includegraphics[scale=0.5]{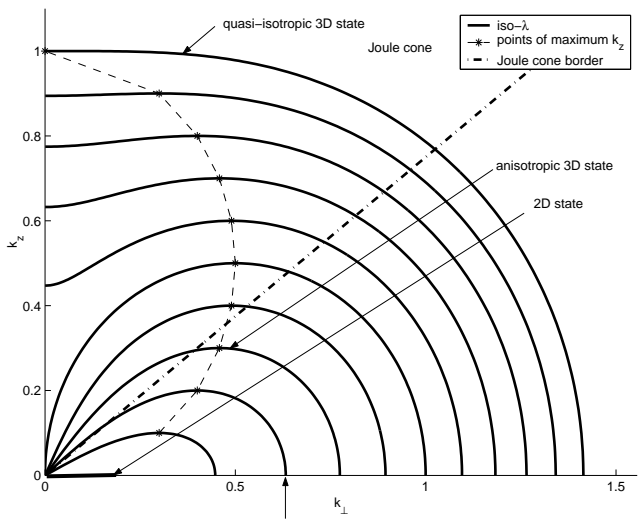}
\caption{Iso-$\lambda$ curves in the plane ($k_\perp,k_z$). One can see the three major types of 
mode distribution: the 2d state corresponds to a set of modes located on the $k_\perp$ axis, the 
strongly anisotropic 3d state exhibits the Joule cone-like shape (the replace of the Joule cone 
has been plotted in the case where all the modes are inside the curve designated by the vertical arrow)
and the quasi-isotropic state is reached when the modes are enclosed inside curves located the furthest 
away from the origin. Axis units are arbitrary.}
\label{fig:cardio}
\end{figure}

%
This provides enough information to visualise the distribution of the $n$ minimal modes:
As shown on figure \ref{fig:cardio}, the 2d manifolds $\lambda=\lambda_m(n)$ are represented 
in the plane $(k_\perp,k_z)$ by a family of curves $G_{Ha,n}$ of equation in 
polar coordinates:
\begin{equation}
G_{Ha,n}:r=\sqrt{-\lambda_m-Ha^2\sin^2\theta}
\label{eq:cardio}
\end{equation}
Note that  $\lambda_m(n)$ is the value which corresponds to the $n^{th}$ mode as the modes are sorted by growing 
 dissipation rate. For $Ha$ fixed, $-\lambda_m(n)$ is thus an increasing function of $n$ and determines uniquely
 the graph $G_{Ha,n}$ when $n$ varies.
 One can then already get a quantitative picture of the set of $n$ minimal modes and distinguish three
 different kinds of sets:\\
1) For  $Ha>1$ fixed, the least dissipative modes are  located on 
the $k_\perp$ axis, and therefore do not depend on $z$ and correspond to a two-dimensional flow independent
of $z$. Indeed,the function $k_\perp\mapsto\-\lambda(k_\perp,k_z)$, where $k_\perp=
\sqrt{k^2_x+k_y^2}$, has a unique absolute minimum for $k_\perp^2=(Ha-k_z)k_z$. The less
dissipative mode is then $(k_\perp,k_z)=(0,1)$, as $(0,0)$ is not permitted (because $\lambda=0$
is not an eigenvalue of $\mathcal D_{Ha}$).
The first 3D mode to appear has to be the least dissipative mode such that $k_z>0$, the variations 
of $\lambda(k_\perp,k_z)$ imply that it is $(Ha-1, 1)$. If $k_{2d}$ is the maximum value of $k_\perp$
among the 2d modes, then
 $(k_{2d},0)$ has to be less dissipative than  $(\sqrt{Ha^2-1}, 1)$, which yields $k_{2d}$ and the 
associated dissipation rate:
\begin{eqnarray}
k_{2d}&=&\sqrt{2Ha}-1
\label{eq:cond2d}\\
\lambda(\sqrt{Ha-1},1)&=&2 Ha
\end{eqnarray}
(If $Ha<1$, the second mode is 3D and is always $(0,1)$).\\
2) The next added modes (by order of growing dissipation rate) spread inside a cardioid 
which is very elongated along the 
$k_\perp$ axis. The flow represented by such modes is therefore highly anisotropic and 
features vortices
stretched along the $z$ direction. Such a distribution of modes matches the well known properties of 
3D turbulent flows under strong magnetic field, for which turbulent modes are located outside the 
so-called Joule cone of axis $k_z$ in the Fourier space (\cite{sm82,alemany79}). More 
precisely, when 
$\sqrt{-\lambda_m}/Ha<1$ (which can only happen for $Ha>1$ for which  $(0,1)$ is not 
the second dissipative mode), the cardioid is located under the line
$\theta=\theta_m$ where $\theta_m=\arcsin\left(\sqrt{-\lambda_m}/Ha\right)$, so that the volume defined by such 
a cardioid elongated along the $k_\perp$ axis matches well a truncation (because $n$ is finite) 
of the space outside the Joule cone.\\
3) Eventually, if  more modes need to be added in order to reach the value of $n$, a value of 
$\lambda_m$ is reached such that $\sqrt{-\lambda_m}/Ha>1$, so that $0<\theta<\pi/2$. In other
words, the cardioid looks more like a quarter of an ellipsoid centred around the origin, the shape 
of which tends toward a quarter circle as the number of modes increases and the Joule cone
degenerates into the $k_z$ axis.
Such a picture describes a nearly isotropic flow, only weakly affected by electromagnetic 
effects. Note that unlike the 2D-3D transition, which takes place for
one specific value of $n$ at $Ha$ fixed, the transition between 3D turbulence with a joule cone
and quasi-isotropic 3d turbulence is smooth. Indeed, for $Ha^2<\lambda_m(n)<2Ha^2$ the $G_{Ha,n}$
 graphs look like some hybrid between a cardioid and an ellipse (see figure \ref{fig:cardio}).\\
In the whole eigenvalue problem,  we have assumed that $n$ was fixed.
However, for a given values of $Ha$ and $Re$, $n$ corresponds to the number of modes for 
which the minimal dissipation compensates the expansion of the initial $n$-volume in the 
phase space due to inertia. At fixed $Ha$, we can see that for low inertia (\textit{i.e.} 
low $n$) the  
flow is two dimensional, whereas for strong inertia, the flow can be close to three dimensional 
isotropic turbulence. Physically, this suggests that the flow corresponding to the estimate 
of the attractor dimension we are
looking for results from a random production of modes by inertia (as the estimate (\ref{eq:trb}) depends
on the number of modes but not on their distribution in the phase space), and a selection of the 
least dissipative modes by the dissipative terms. Of course, this assumes that 
the estimate (\ref{eq:trb}) for the 
expansion rate due to inertial effects is realistic, at least with regard to its dependency on $n$.
At this point, it is important to recall that the minimal modes of the dissipation operator found 
here are not solution of the Navier-Stokes equations. However, they turn out to 
exhibit a physical behaviour which matches qualitatively what is heuristically known from turbulent 
MHD flows. This suggests that expansions of solutions of the Navier-Stokes equations over the base of 
minimal eigenmodes of the dissipation might be suitable to calculate turbulent flows, 
all the more as these modes already satisfy the boundary conditions.\\
We shall now compute recursively the set of $n$ least dissipative modes in the
 discrete space of Fourier coefficients and find the related upper bound 
for the attractor dimension. Note that as we 
actually construct an $n$-dimensional set of modes which achieves the  maximum magnitude 
of the trace of $\mathcal D_{Ha}\mathbf P_n$, the upper bound for the modulus of this trace
actually \textit{is} the maximum (keeping in mind that the trace of the dissipation is negative). 

\subsection{trace of the dissipation operator and attractor dimension}
\label{sec:numeric}
We shall now calculate the trace of $\mathcal D_{Ha}\mathbf P_n$ associated with the least dissipative modes 
as a function of $n$ and $Ha$, then using (\ref{eq:trb}), we express the estimate for 
the upper bound of the attractor dimension as a function of $Ha$ and $Re$ by searching the 
value of $n$ which annihilates the trace of the evolution operator, as explained in section 
\ref{sec:method}. The trace 
of $\mathcal D_{Ha}\mathbf P_n $ is calculated nearly exactly using a computer 
(the only error is due do 
to truncation after the $17^{th}$ digits of real numbers which occurs in our program) by adding 
up the dissipation
rates along the sequence of modes sorted by increasing values of $-\lambda$. The method is described in 
appendix \ref{ap:numeric}.
\begin{figure}
\begin{center}
\includegraphics[scale=0.5]{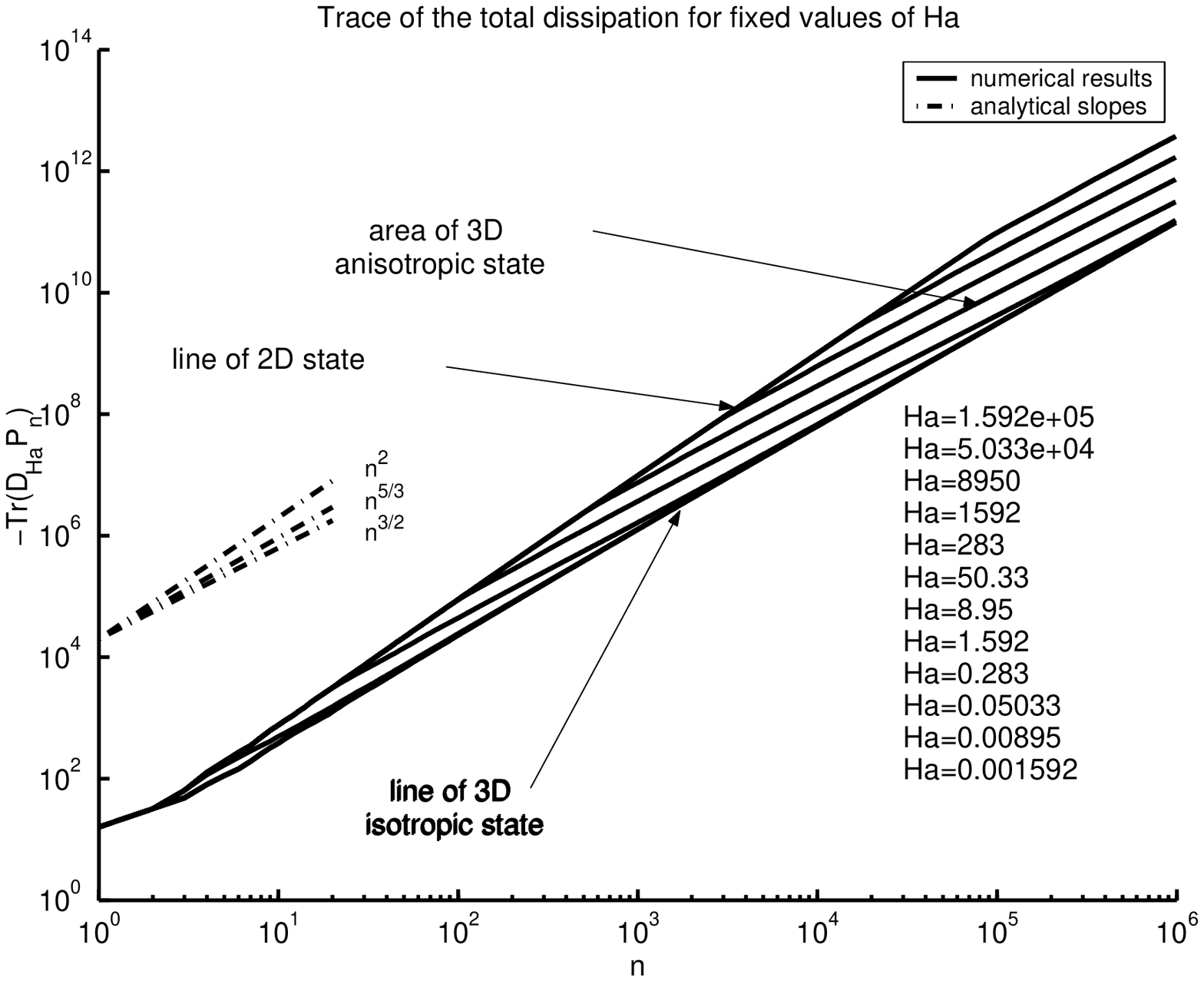}
\includegraphics[scale=0.5]{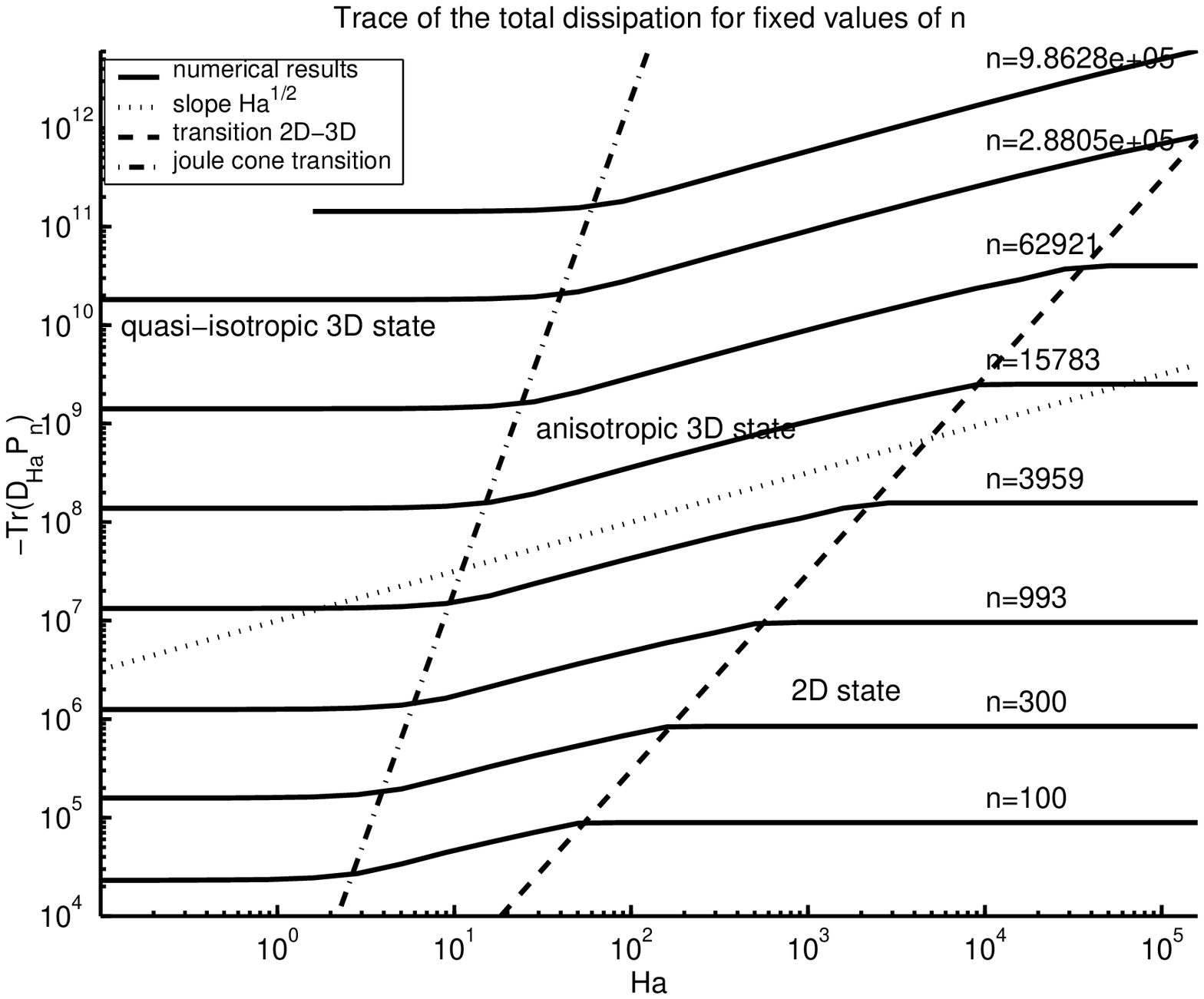}
\caption{Trace of the dissipation operator as a function of $n$ for fixed $Ha$
(left).
Each iso-$n$ curve exhibits successively $n^2$ (2d modes), $n^{3/2}$ (3d anisotropic set of modes)  and $n^{5/3}$ slopes (quasi-isotropic set of modes).
The higher the value of $Ha$, the later the transitions occur.The curves 
corresponding to the 5 lowest values of $Ha$ are not distinguishable.
Trace of the dissipation operator as a function of $Ha$ for fixed $n$
(right). The three different kinds of sets of modes appear (quasi-isotropic, for
low $Ha$, 2d for high $Ha$).}
\label{fig:trace}
\end{center}
\end{figure}
\begin{figure}
\begin{center}
\includegraphics[scale=0.5]{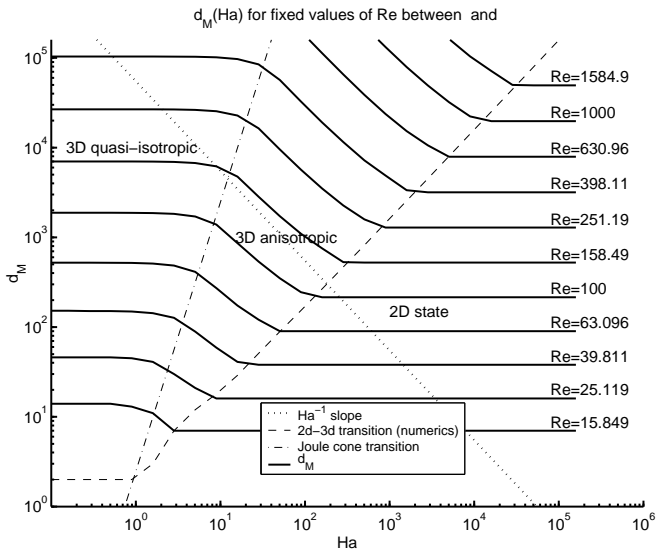}
\caption{Attractor dimension as a function of  
$Ha$ for fixed $Re$}
\label{fig:dimat}
\end{center}
\end{figure}
\begin{figure}
\begin{center}
\includegraphics[scale=0.5]{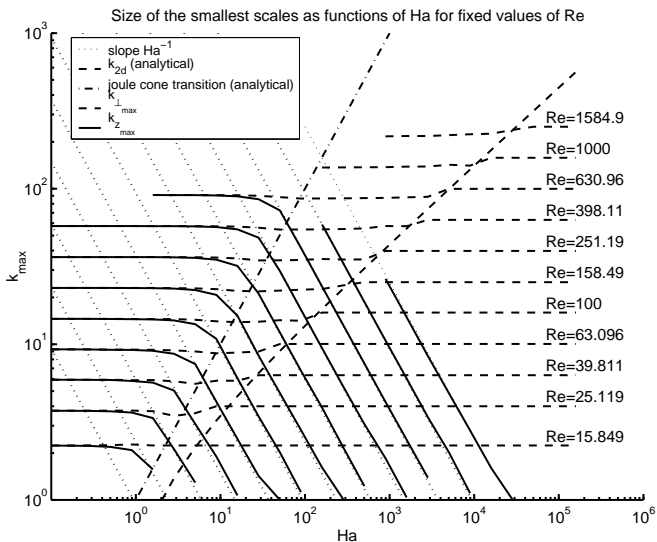}
\caption{Size of the smallest scales in the direction of the magnetic field 
and perpendicular to the magnetic field. Dotted: $Ha-1$ slope, dashed: $k_{2d}$ analytical,
dash-dot: Joule cone transition (analytical), dashed (quasi-horizontal lines): $k_{\perp_max}$ (numeric)
solid: $k_{z_{max}}$ (numeric)}
\label{fig:sscales}
\end{center}
\end{figure}
The graphs
of $\Tr(\mathcal D_{Ha}\mathbf P_n)$ and the estimate for the attractor dimension $d_M(Re,Ha)$ are reported 
respectively on figures \ref{fig:trace} and \ref{fig:dimat}. For now, let us put the emphasis
on the curves on figure \ref{fig:dimat} which show the variations of the attractor dimension estimate with regard to the 
variable $Ha$, for different fixed values of the Reynolds number. Apart for very low values 
of the Reynolds number (of the order of unity,
which does not relate to the usual picture of turbulent flows), each curve clearly exhibits 
three distinct regions corresponding to three different ranges of Hartmann numbers. Let us  
follow a given curve from $Ha=0$ to high values of $Ha$. Physically, this would correspond
to looking at a turbulent flow and increasing the applied magnetic field in a quasi-static
way:\\
1) We first encounter a region where the attractor dimension is nearly constant when the
Hartmann number increases. This region describes a flow under weak magnetic field, for which
the dissipation is essentially due to viscosity, and therefore does not depend on the magnetic 
field. The flow is in a state of 3d quasi-isotropic turbulence and the modes are spread within
a nearly circular  (or radius $k_{\perp_m}\sim k_{z_m}$) region of the $(k_\perp,k_z)$ 
plane.\\
2) For values of $Ha$ above one, the attractor dimension decreases approximately as $Ha^{-1}$. 
Indeed,
for $Ha \sim 1$ viscous and Joule dissipation are of the same order of magnitude so that the 
overall dissipation is stronger than in the hydrodynamic case. Therefore as $Re$ is fixed, 
fewer modes are needed to reach a dissipation which balances the expansion rate $nRe$ due to 
inertia. Equivalently, turbulence becomes more and more anisotropic as vortices are 
stretched in the direction of the magnetic field, so if one interprets $d_M$ as the number of
vortices in the domain, as in \cite{constantin85_jfm}, fewer of these long vortices are needed to fill 
the $2\pi L\times 2\pi L \times 2\pi L$ box, 
so that the number of degrees of freedom decreases. This second case corresponds to a set of 
modes defined by the elongated cardioid of section \ref{sec:eigenval}\\
Eventually, for even higher values of the 
magnetic field, one reaches a region where again, the estimate found for the 
attractor dimension does not depend on $Ha$. One can see from figure \ref{fig:dimat} that this 
happens when the flow undergoes a transition between 3d and 2d state, or in other words, when 
 even the smallest vortex reaches the size of the box in the $z$ direction ($k_{z_m}< 1$). 
 The flow then becomes two-dimensional and 
looks like "rows" of columnar vortices (modes are on the $k_\perp$ axis according to the
description of section \ref{sec:eigenval}).
As there is no more velocity variation along the magnetic
field lines, no current loops are present in the flow so that the Lorentz force falls to $0$
 (looking at (\ref{eq:roberts}), $\partial^2_{zz}\mathbf u = 0$ implies 
 $\mathbf j \times \mathbf B =0$). The attractor dimension does not depend on $Ha$ anymore but
 should match estimates found for two-dimensional turbulence. It does not turn out to be  
 the case but we shall leave this point for more thorough discussion in section 
 \ref{sec:heuristics}.\\ 
Up to now, we have found a set of modes which corresponds to the actual minimum dissipation and 
which, under the assumption of finite dissipation (\ref{eq:eps}), returns an upper bound for 
the attractor dimension of turbulent MHD flow, without any restrictions on the 
values of $Ha$ and $Re$. What is more striking is that although these modes are not solution 
of the Navier-Stokes equations themselves, their distribution in the Fourier 
space seems to match physical observations for such turbulent 
flows. To find 
out to what extend this is the case, we shall now 
derive some analytical approximations of the "exact" results found in this section 
and compare them to broadly accepted results derived from heuristic arguments.
\section{Asymptotic results and comparison with heuristics}
\label{sec:heuristics}
\subsection{integral formulation of the eigenvalue problem}
\label{sec:int}
We go back to the point where the set of $n$ less dissipative modes is calculated, at the end
of section \ref{sec:distrib}, and we aim at finding some analytical approximations for the 
results obtained numerically in section \ref{sec:numeric}, should it be at the price of 
working only in asymptotic regimes of the flow parameters. At this stage, the problem of 
finding the set of $n$ eigenmodes which minimise the dissipation can be mathematically 
formulated as follows: for given $n$ and $Ha$, we look for the set of $n$ points 
$(\mathbf{k}_i)_{i=1..n}$ in $\mathbb N^3-(0,0,0)$ which achieves the minimum of the functional:
\begin{equation}
\Tr(\mathcal D_{Ha}\mathbf P_{8n})=2\sum_{i\in\{1..n\}}-\lambda(\mathbf k_i)
\label{eq:trd_n}
\end{equation}
The study of this set of modes in section \ref{sec:eigenval} has shown that the $(\mathbf k_i)_{i=1..n}$
are located inside the volume $V_{\lambda_m}$ located "under" the manifold of equation 
$\lambda(k_x,k_y,k_z)=\lambda_m$,
which defines it uniquely\footnote{see note at the beginning of section \ref{sec:distrib}}.
The problem then comes down to expressing
$\lambda_m$ as a function of $n$ and $Ha$. To this end, we notice that the dimension of the attractor 
associated with a turbulent flow is an enormous number, for which the sums (such as the one in
(\ref{eq:trd_n})) can be safely replaced by integrals over the continuous Fourier space. 
Under this approximation, the fact the the modes are on a discrete set implies that each of them fills a 
unit-volume in the Fourier space,
so that the volume contained under the manifold $\lambda(x,y,z)=\lambda_m$ should be $n/8$:
\begin{equation}
16\int_{V_{\lambda_m}}dk_xdk_ydk_z=n
\label{eq:nmodes}
\end{equation}
The trace of $\mathcal D_{Ha}\mathbf P_n$ similarly expresses as:
\begin{equation}
\Tr(\mathcal D_{Ha}\mathbf P_n)=16\int_{V_{\lambda_m}}\lambda(k_x,k_y,k_z)dk_xdk_ydk_z
\label{eq:trdint}
\end{equation}
Equations (\ref{eq:nmodes}) and (\ref{eq:trdint}) allow to derive both $\lambda_m$ and 
$\Tr(\mathcal D_{Ha}\mathbf P_n)$ as functions of $n$ and $Ha$ only. This can be done analytically 
all the way for through each of the types of minimal set of $n$ modes found in section \ref{sec:eigenval}.
The next two sections are devoted to this task, as well as to comparing the obtained results 
with heuristic considerations on MHD Turbulence.
\subsection{Anisotropic turbulence under strong magnetic field}
\label{sec:strongb}
\subsubsection{Analytical estimates}
Let us first tackle the case where the modes are located within an elongated cardioid, \textit{i.e.}
$Ha<-\lambda_m<Ha^2$, which corresponds to a 3d anisotropic flow with dominant electromagnetic effects 
(see section \ref{sec:numeric}).
After integration in cylindrical coordinates for $0<\theta<\theta_m$, (\ref{eq:nmodes}) and 
(\ref{eq:trdint}) respectively take the form:
\begin{eqnarray}
\frac{n}{Ha^3}=\frac{\pi^2}{2}\sin^4 \theta_m\\
\label{eq:n_theta}
\frac{\Tr(\mathcal D_{Ha}\mathbf P_n)}{Ha^5}=-\frac{\pi^2}{3}\sin^6\theta_m
\label{eq:tr_theta}
\end{eqnarray}
Let us recall that $\theta_m$ is defined by $\sin\theta_m=\frac{\sqrt{-\lambda_m}}{Ha}$, hence equation
(\ref{eq:n_theta}) allows us to express $\lambda_m$ as a function of $n$:
\begin{equation}
\lambda_m=\frac{\sqrt{2}}{\pi}Ha^{1/2}n^{1/2}
\end{equation}
and equation (\ref{eq:tr_theta})
allows then to express the trace of the dissipation in terms of $n$:
\begin{eqnarray}
\Tr(\mathcal D_{Ha}\mathbf P_n)=\frac{2\sqrt{2}}{3\pi}n^{3/2}Ha^{1/2}\\
\sin\theta_m=\sqrt{\frac{2}{\pi}}n^{1/4}Ha^{-3/4}=\sqrt{\frac{-\lambda_m}{Ha^2}}
\label{eq:lambdam}
\end{eqnarray}
The value of $n$ for which the trace of the total evolution operator is zero (\textit{i.e.} 
$\Tr((\mathcal D_{Ha}+ \mathcal B(.,\mathbf u)\mathbf P_n))=0)$ is an upper bound for the attractor 
dimension, so using (\ref{eq:trb}):
\begin{equation}
d_M \leq \frac{9\pi^2}{32}\frac{Re^4}{Ha} 
\label{eq:dimat1}
\end{equation}
The geometrical shape of the cardioid which defines the set of $n$ minimal modes 
yield the maximum values reached by $k_\perp$ and $k_z$ respectively:
\begin{eqnarray}
k_{\perp_m}&=& \sqrt{-\lambda_m}=\frac{2^{1/4}}{\pi^{1/2}}n^{1/4}Ha^{1/4}
\label{eq:kpmax0}\\
k_{z_m}&=&-\frac{\lambda_m}{2Ha}=\frac{1}{\pi\sqrt{2}}n^{1/2}Ha^{-1/2}
\label{eq:kzmax0}
\end{eqnarray}
The bounds for the size of the small scales are obtained  by replacing $n$ by $d_M$ 
(\ref{eq:kpmax0}) and (\ref{eq:kzmax0}) respectively:
\begin{eqnarray}
k_{\perp_m}\leq {\frac{\sqrt{3}}{2}} Re
\label{eq:kpmd1}\\
k_{z_m} \leq \frac{3}{8}\frac{Re^2}{Ha}
\label{eq:kzmd1}
\end{eqnarray}
Graphs of the relations (\ref{eq:dimat1}), (\ref{eq:kpmd1}) and (\ref{eq:kzmd1}) are plotted 
on figures \ref{fig:dimat} and \ref{fig:sscales} respectively, along with the numerical results of 
section \ref{sec:numeric} and bring confirmation that the 
discrete Fourier space can be accurately approached by a continuum.\\
\subsubsection{Heuristics on MHD turbulence of Kolmogorov type under strong field}
Now, it is  worth 
underlining again that these results are exact, and come exclusively from the 
mathematical properties of the Navier-Stokes equations, without the involvement of any physical 
approximation. There is therefore considerable interest in comparing  them with orders of 
magnitude obtained from heuristic considerations. Let us recall how the smallest scales can 
be obtained in a more physical manner: in a 3D periodic flow where Joule dissipation is stronger than
viscosity except at small scales ($Ha>>1$), it is usual to  consider that a vortex in the inertial 
range (\textit{i.e}
 not destroyed by viscosity) results from a balance between inertial and Lorentz forces, which 
 implies:
\begin{equation}
\frac{k_z}{k_\perp} \sim \left(\frac{\sigma B^2 L}{\rho k_\perp U_v}\right)^{-1/2}
\label{eq:anis}
\end{equation}
Moreover, one usually assumes that anisotropy remains the same at all scales \cite{alemany79}, 
over the inertial range. Under this assumption, (\ref{eq:anis}) implies  $U_v(k_\perp)=U_0k_\perp^{-1}$,
where $U_0$ stands for a typical large scale velocity. 
This is usually expressed in terms of the energy spectrum as:
\begin{equation}
E(k_\perp)\sim k_\perp^{-1} U_v^2(k_\perp) \sim U_0^2k_{\perp}^{-3}
\label{eq:k-3}
\end{equation}
and allows to rewrite (\ref{eq:anis}) as:
\begin{equation}
\frac{k_z}{k_\perp} \sim \frac{Re_0^{1/2}}{Ha}=N^{-1/2}
\label{eq:anis_h}
\end{equation}
$Re_0$ is a Reynolds number scaled on $U_0$ and $L$, the ratio $\frac{Ha^2}{Re_0}=N$ is the 
corresponding interaction parameter. Eventually, the small scales  are 
heuristically defined as the smallest possible structures of the inertial range which are not 
destroyed by viscosity, which means that they results from a balance between inertia and viscosity.
This yields:
\begin{equation}
\frac{k_{z_m}}{k_{\perp_m}^2} \sim Ha^{-1}.
\label{eq:sscales}
\end{equation}
Now combining (\ref{eq:anis_h}) and (\ref{eq:sscales}) yields:
\begin{eqnarray}
k_{\perp_{max}} \sim Re^\frac{1}{2}\\
k_{z_{max}} \sim \frac{Re}{Ha}
\label{eq:kzmp1_h}
\end{eqnarray}
from which the number of degrees of freedom of the flow can be estimated by counting the number of vortices in 
the of size $L/k_\perp\times L/k_\perp \times  L/k_z$ in a $L\times L\times L\times$ box:
\begin{equation}
N_f \sim k_\perp^2k_z \sim \frac{Re^2}{Ha}
\label{eq:nf_h}
\end{equation}
When comparing $N_f$ to $d_M$ and the heuristic small scales to (\ref{eq:kpmd1}) and (\ref{eq:kzmd1}),
we see that our mathematical estimates are loose when compared to heuristics because they exhibit 
a higher exponent of the Reynolds 
number than the heuristic relations. However, exponents of the Hartmann number match, which suggests
that the mathematical study actually captures well the electromagnetic effects in turbulence. 
This is confirmed by the fact that the sizes of the smallest scales for a given number of modes $n$
are exactly  matched by heuristic results presented in this  sections ((\ref{eq:kpmax0}) and 
(\ref{eq:kzmax0}) can indeed be recovered
 from (\ref{eq:sscales}) and (\ref{eq:nf_h}), considering $n \sim N_f$ vortices in a box).
This, together with the fact that our estimate for the trace of the dissipation corresponds to an
achieved extremum suggests that the latter is optimal.
Besides, if one considers $d_M \sim N_f \sim \frac{Re^2}{Ha}$ as the order of magnitude expected 
for the attractor dimension, then one should expect the trace of the operator defined by inertial
terms to be of the order of $|\Tr(\mathcal B(.,\mathbf u))| \sim n Re$ for $d_M\sim \frac{Re^2}{Ha}$
to be solution of $\Tr((\mathcal D_{Ha}+\mathcal B(.,\mathbf u))\mathbf P_n)=0$. This suggests that 
the exponent of $n$ is optimal in  (\ref{eq:trb}), whereas the exponent of $Re$ is somewhat too
high to match heuristic results valid for MHD turbulence of Kolmogorov type (\textit{i.e.} with
an established turbulent spectrum). Note that  $d_M\sim \frac{Re^2}{Ha}$ and (\ref{eq:lambdam}) 
yield $\sin\theta_m\sim \sqrt{Re}/Ha$ which matches the prediction of \cite{sm82} for the Joule
cone angle, whereas the rigorous estimate (\ref{eq:dimat1}) again yields an overestimated exponent 
for $Re$ but the right one for $Ha$.

\subsection{3D turbulence under weak magnetic field ($Ha<<1$)}
\label{sec:weakb}
\subsubsection{Analytical estimates}
Let us now investigate the case where $-\lambda_m/Ha^2>1$ (and $\theta_m=\pi/2$) which relates to 
weakly anisotropic turbulence,
as mentioned in section \ref{sec:eigenval}. After integration in cylindrical coordinates, for 
$0<\theta<\pi/2$, (\ref{eq:nmodes}) and (\ref{eq:trdint}) rewrite respectively:
\begin{eqnarray}
\frac{n}{Ha^3} = 
\nonumber \\
\frac{\pi}{6}\left(5l\sqrt{l-1}
-2\sqrt{l-1}
+3l^2\arctan\left(\frac{1}{\sqrt{l-1}}\right)\right)
\label{eq:n_modes_iso}
\\
\frac{\Tr(\mathcal D_{Ha} \mathbf P_n)}{\pi Ha^5}=
\nonumber \\
\frac{14}{15}\sqrt{l-1}
-\frac{4}{45}l\sqrt{l-1}
-\frac{8}{45}\sqrt{l-1}+\frac{2}{3}l^3\arctan{\frac{1}{\sqrt{l-1}}},
\label{eq:trd_iso}
\end{eqnarray}
with $l=-\frac{\lambda_m}{Ha^2}$. It is here more difficult to the extract analytical expression for $\Tr (\mathcal D_{Ha}\mathbf P_n)$ as a function 
of $n$ and $Ha$. However, equations can be expanded in powers of $l$ in the limit $l\rightarrow \infty$.
This corresponds to a flow where inertia is large compared to inertial effects. Keeping only the terms in
$l^{3/2}$ and $l^{1/2}$ in the expansion of (\ref{eq:n_modes_iso}), $l$ can be expressed as a function of $n/Ha^3$. 
Assuming this latter parameter is large as well and keeping the leading two terms yields:
\begin{equation}
-\lambda_m\simeq \frac{1}{4}\frac{2}{\pi}n^{2/3}+\frac{1}{3}Ha^2
\label{eq:lambdam2}
\end{equation}
also, keeping the two leading powers of $l$ in the expansion of (\ref{eq:trd_iso}) and using (\ref{eq:lambdam2})
yields:
\begin{equation}
\Tr(\mathcal D_{Ha}\mathbf P_n)\simeq
\frac{3}{10}\left(\frac{6}{\pi}\right)^\frac{2}{3}n^\frac{5}{3}+\frac{2}{3}Ha^2n
\end{equation}
As in the case of strong fields, the upper bound for the attractor dimension if obtained by looking 
for the value of $n$ which annihilates the trace of the evolution operator:
\begin{equation}
d_M \leq \frac{15^{3/2}}{162}\sqrt{5} Re^3 \left(1-\frac{4}{3}\frac{Ha^2}{Re^2} \right)^{3/2}
\label{eq:dimat2}
\end{equation}
In a quasi-isotropic flow the set of minimal modes spread in an ellipsoid-like volume of the phase space
so that The maximum values of $k_\perp$ and $k_z$ 
are obtained for $\theta=0$ and $\theta=\pi/2$ respectively:
\begin{eqnarray}
k_{\perp_m}&=&\sqrt{-\lambda_m}
\label{eq:kpmax}\\
k_{z_m}&=&\sqrt{-\lambda_m-Ha^2}
\label{eq:kzmax}
\end{eqnarray}
The bounds for the size of the small scales are obtained  by replacing $n$ by $d_M$
in (\ref{eq:lambdam2}) and using (\ref{eq:kpmax}) and (\ref{eq:kzmax}) respectively:
\begin{eqnarray}
k_{\perp_m}=\frac{\sqrt{5}}{2\times3^{5/6}}Re\left(1-\frac{4}{3}\frac{Ha^2}{Re^2}(1-\frac{3^{2/3}}{5}) \right)^{1/2}
\label{eq:kpmd2}\\
k_{z_m}=\frac{\sqrt{5}}{2\times3^{5/6}}Re\left(1-\frac{4}{3}\frac{Ha^2}{Re^2}(1+2\frac{3^{2/3}}{5}) \right)^{1/2}
\label{eq:kzmd2}
\end{eqnarray}
These final results on the dimension of the attractor and associated small scales match well the 
properties of the flow put in light by the numerical results of section \ref{sec:numeric}: in 
the limit of low $Ha$, both $d_M$, $k_\perp$ and $k_{z_m}$ are weakly dependent on electromagnetic
effects. The flow is indeed almost isotropic apart from  a slight vortex elongation in the $z$ 
direction. Also, the upper
 bound for the attractor dimension in classical 3D turbulence (\ref{eq:dim3d}) is recovered for
 $Ha=0$. Note that when $Ha$ is progressively increased from $0$,  the small scales initially grow both in 
 the direction orthogonal to the field and in the direction of the field. However, the growth is more important 
 in the direction of the field which results in an early anisotropy. It can be seen from figure (\ref{fig:sscales})
 that when $Ha$ is increased up to a value where Lorentz dissipation is more important than viscous 
 dissipation, the length scale in the direction orthogonal to the field saturates at the value found in 
 (\ref{eq:kpmd1}) whereas the length scale in the direction of the field continues to grow until it reaches the 
 typical macroscopic length scale.\\
\subsubsection{Heuristic considerations in quasi-isotropic MHD turbulence of Kolmogorov type}
 %
%
%
%
When eletromagnetic effects are small compared to inertia, the turbulence is almost 3D 
isotropic, so one expects the size of the small scales to be close to the value 
$k_m^{-1}\sim Re^{-3/4}$ obtained from the K41 theory \cite{k41}. Indeed, \cite{alemany79} 
proposes some heuristic estimates which suggest that the size of the small scales if of
 this order of magnitude, and tends to slightly increase under the effects of small 
 eletromagnetic effects. To this regard, the mathematical estimates  
 (\ref{eq:dimat2}), (\ref{eq:kpmd2}) and (\ref{eq:kzmd2}) again
exhibit higher exponents of $Re$ than heuristic results which suggests inertial effects are 
overestimated. Indeed, as in the case of strong magnetic fields, our estimate for the trace of 
the dissipation is optimal, so if one is to trust heuristic values of the small scales, then
a better estimate for the trace of the inertial terms is expected to be of the order of $n Re^{3/2}$,
which is smaller than (\ref{eq:trb}). As for strong fields, the exponent of $n$ in
(\ref{eq:trb}) seems to be optimal whereas the exponent of $Re$ is overestimated. It is however 
remarkable that the exponent expected from heuristic considerations for a weak field is different 
than the one which would be expected for strong fields. This suggests  finding a better estimate than 
(\ref{eq:trb}) for the 
trace of the inertial terms would need to account for the mode distribution in the Fourier space.
\subsection{The 2D case}
For two-dimensional flows (\textit{i.e} $\lambda_m\leq 2Ha$), the motion equations reduce to 2d Navier-Stokes 
equations without magnetic
field, as the Lorentz force falls to zero. The dissipation operator is a simpler two-dimensional 
Laplacian operator, for which the trace of any $n$ dimensional subset of the phase space is bounded
by (see for instance \cite{doering95}, or using the approximation of a continuous Fourier space 
as all along this section):
\begin{equation}
\Tr(\mathcal D_{Ha} \mathbf P_n) \leq \frac{n^2}{2\pi}
\label{eq:trd2d}
\end{equation}
which, together with (\ref{eq:trb}) leads to an upper bound for the attractor dimension:
\begin{equation}
d_M \leq \frac{\pi}{8}Re^2
\label{eq:dimat2d_3d}
\end{equation}
The estimate (\ref{eq:dim2d}) presented in \cite{doering95} for $d_M$ is based on an accurate estimate for the
2D inertial terms of the order of $n^{1/2}\mathcal G(1+\ln n)^{3/4}$. Although it is difficult to compare 
the Grashof number $\mathcal G= \|\mathbf f\|_2/\nu$ (where $\|\mathbf f\|_2$ stands for the $\mathcal L^2$ norm
of the dimensional forcing $\mathbf f$) to the Reynolds number, one can be sure that the estimate (\ref{eq:dimat2d_3d})
is rather bad in the 2D case, as it features a much higher exponent of $n$ than the estimate from \cite{constantin88}.
This again supports the idea that a sharp estimate for the inertial terms must account for the modes distribution
(the estimate by \cite{constantin88} is derived from the 2d assumption whereas (\ref{eq:trb}) is a generic 3d result)
However, as both (\ref{eq:dim2d}) and our estimate for the transition between 2D and 3D state (\ref{eq:cond2d})
are consistent with heuristics one can expect them to yield a realistic transition curve in the $(\mathcal G,Ha)$ plane.
The latter is obtained by noticing that the $iso-\lambda(k_x,k_z)$  curves for $k_z=0$ are circles centred on the
origin so that under the approximation of a continuous Fourier space, the number of 2D modes is 
$n_{2d}=2\pi k_{2d}^2$ (the factor 2 is due to the fact that eigenspaces are of dimension 2).
Setting $n_{2d}$ to the value of $d_{2d}$ given by (\ref{eq:dim2d}) and using (\ref{eq:cond2d}) yields the transition
curve:
\begin{equation}
(\sqrt{2Ha}-1)^2=\frac{c_1}{2\pi}\mathcal G^{2/3}(1+\ln\mathcal G)^{1/3}
\label{eq:transition}
\end{equation}
The fact that the transition is expressed using $Ha$ and $\mathcal G$ makes it all the more applicable 
to experimental
configurations as it only depends on the control parameters, unlike the Reynolds
 number which involves a velocity which can be hard to define and sometimes to 
measure.

\section{Concluding remarks}
We have found a rigorous upper bound for the attractor dimension in low-$Rm$ MHD turbulence, which is valid for all values of $Ha$ and $Re$, and relies solely on the 
Navier-Stokes equations. This bound is obtained for the set of modes which 
achieves the minimum of the total dissipation (viscous and Joule). This 
particular set of modes exhibits most of the well known features of MHD 
turbulence: quasi-isotropic turbulence close to hydrodynamic turbulence for weak
electromagnetic force, strongly anisotropic state (modes inside the Joule cone in
 the Fourier space) when Joule dissipation is of the order of viscous dissipation, 
 and two-dimensional state when the Joule dissipation is dominant. The related
 estimates for the small scales and Joule cone angle show the same dependence 
 on $Ha$ as their heuristic counterpart. However, because the estimate we use
 for the inertial terms is not optimal, the exponent of $Re$ in the final 
 attractor dimension is higher than predicted by heuristic considerations. 
 It is noteworthy that this discrepancy to the heuristics is not the same for the three
 different kinds of turbulence pointed out above , which are characterised by 
 three very different 
 modes distributions (3d isotropic, 3d anisotropic with Joule cone, and 2d 
 isotropic). This  suggests that a better estimate for the inertial terms can 
 be obtained by accounting for the mode distribution in the Fourier space.\\
 However, The result found for the transition (\ref{eq:transition}) between 2D 
 to 3D  turbulence does not suffer form this limitation on the estimation 
 of the inertial terms as it is derived from estimates for inertia and 
 dissipation which both match heuristic results. This simple analytical 
 result now needs testing against experiment.\\
 The other possible improvement to the results found here has to to with 
 the periodical conditions in space. Indeed, in laboratory experiments, as well
as industrial setups, the 2d state is achieved when the flow is confined between 
two walls perpendicular to a strong magnetic field, so that the dissipation along
these walls (in the Hartman boundary layer) is often the main factor which 
determines the whole flow \cite{psm00}. This makes the 2D state obtained 
under 3d periodical conditions rather unphysical. A way to improve this result
 would be to carry out the same study as presented here with walls in $z=0$
  and $z=1$. Unfortunately, this will be at the expense of a more complex calculation for 
  which no analytical estimate can be derived.\\
  	Eventually, the fact that the least dissipative modes already incorporates
many properties of MHD turbulence encourages us to 
investigate their ability to reproduce the energetic properties of MHD 
turbulence such as the the $k^{-3}$ spectrum observed in the 3d anisotropic 
regime \cite{alemany79}. Also, it may be possible to reproduce
the main properties of the flow using a reduced set of these modes 
in a numerical model. Indeed, because of the strong anisotropy which characterises 
Low-$Rm$ MHD turbulence under strong magnetic field, it is in principle possible 
to fully describe this class of 
flow using $\frac{Re^2}{Ha}$ taken from the set of least dissipative modes. This 
represents a much smaller set of modes than the $Re^2$  modes obtained by taking 
all the fourrier modes of wavenumber smaller than $k_{\perp_m}$.\\

The authors would like to aknowledge financial support from the Leverhulme Trust, under 
Grant $F/09 452/A$.
 
\appendix
\section{Numerical calculation of the least dissipative modes}
\label{ap:numeric}
\begin{figure}
\centering
\includegraphics[scale=0.5]{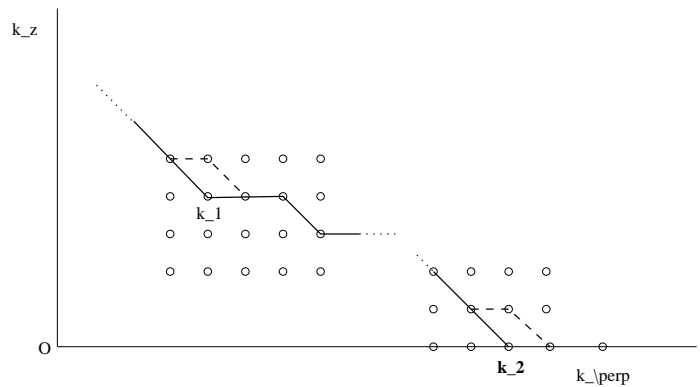}
\caption{Recursive process to find the $n+1^{st}$ modes knowing the first $n$ first (located below 
the solid line, excluding the points on the line). The point $\mathbf k_{n+1}$ which yields the 
minimal dissipation
outside of the set made up with the first $n$ modes is found among the points of the solid line. Once it 
is found by looking at all the values of $-\lambda$ for these points , the $n+2^{nd}$ mode is searched among the 
points of the dashed line, obtained by modifying the solid line so that it "surrounds" $\mathbf k_{n+1}$. Two
distinct examples are given: $\mathbf k_{n+1}=\mathbf k_1$ \textit{or} $\mathbf k_{n+1}=\mathbf k_2$. }
\label{fig:algo}
\end{figure}
The sequence of $n$ eigenmodes of the dissipation operator is calculated recursively, in
growing  order of the eigenvalues' modulus (which represents the dissipation rate of 
the associated eigenmodes). We start from the 
less dissipative mode $(1,0)$ which corresponds to the eigenvalue
of smallest module $\lambda_1=-1$ (or smallest dissipation rate). As
$(k_\perp,k_z)\mapsto-\lambda(k_\perp,k_z)$ has a unique absolute minimum, 
at $(0,1)$ the value of $-\lambda(k_\perp,k_z)$ increases 
along any direction originating from this minimum. 
The following less dissipative values (along the sequence of modes sorted by growing dissipation rate) 
are to be sought in the closest possible vicinity of this minimum (bearing in mind 
that both $k_\perp$ and
$k_z$ span a discrete set of positive values). The second  $\mathbf k$  
is then found by looking for the one which  yields the smallest 
 value of $-\lambda(k_\perp,k_z)$ among the points "surrounding" the minimum. The process is iterated, replacing the point selected 
 from the 
 previous step in the surrounding curve by the set of points surrounding it 
 (and which are not already in the  set of minimal modes) as shown on figure \ref{fig:algo}
 .  Note that this algorithm requires to know the sequence of values of
 $k_\perp$. The latter is calculated using the same process, applied to the function 
 $(k_x,k_y)\mapsto k_x^2+k_y^2$.\\
 It is straightforward to extract the value of $n$ which corresponds to the first 3d minimal 
 mode : this
 gives the 2d-3d transition curve in the $(n,Ha)$ plane. In order to save some calculation time, the 
 attractor's dimension is actually worked out at every added mode: indeed, as the estimate is 
 obtained by writing that the expansion of the $n$-volume in the phase space is the same as the 
 contraction induced by the dissipation, once the maximum trace of the dissipation operator
 is obtained for a given $n$, we calculate the value of $Re$ for which $n$ is an upper 
 bound for the  attractor dimension by:
\begin{equation}
Re=\sqrt\frac{-\Tr(D_{Ha}\mathbf P_n)}{n}
\label{eq:renum}
\end{equation}
The process is iterated using a short program written in the MATLAB environnement.

\begin{thebibliography}{10}

\bibitem{k41}
Kolmogorov A, N.
\newblock local structure of turbulence in an incompressible fluid at very high
  reynolds numbers.
\newblock {\em Dokl. Akad. Nauk. SSSR}, 30:299--303, 1941.

\bibitem{alemany79}
A.~Alemany, R.~Moreau, P.~Sulem, and U.~Frish.
\newblock Influence of an external magnetic field on homogeneous {MHD}
  turbulence.
\newblock {\em Journal de M\'ecanique}, 18(2):277--313, 1979.

\bibitem{constantin87}
P.~constantin.
\newblock Collective l-infinity estimates for families of functions with
  orthonormal derivatives.
\newblock {\em Indiana univ. Math. J.}, 36:603--615, 1987.

\bibitem{constantin85_cpam}
P.~Constantin and C.~Foias.
\newblock Global lyapounov exponents , kaplan-yorke formulas anthe dimension of
  the 2d navier-stokes equation.
\newblock {\em Comm. Pure. Appl. Math.}, 38:1--27, 1985.

\bibitem{constantin85_ams}
P.~Constantin, C.~Foias, O.P. Mannley, and R.~Temam.
\newblock attractors representing turbulent flows.
\newblock {\em Mem. Am. Math. Soc.}, 53,314, 1985.

\bibitem{constantin85_jfm}
P.~Constantin, C.~Foias, O.P. Mannley, and R.~Temam.
\newblock determining modes and fractal dimension of turbulent flows.
\newblock {\em J. Fluid. Mech.}, 150:427--440, 1985.

\bibitem{constantin88}
P.~Constantin, C.~Foias, and R.~Temam.
\newblock on the dimension of the attractors in 2d turbulence.
\newblock {\em physica D}, 30:284--296, 1988.

\bibitem{dav97}
P.~A. Davidson.
\newblock The role of angular momentum in the magnetic damping of turbulence.
\newblock {\em J. Fluid. Mech.}, 336:123--150, 1997.

\bibitem{davidson01}
P.A. Davidson.
\newblock {\em An introduction to magnetohydrodynamics}.
\newblock Cambridge University Press, 2001.

\bibitem{doering95}
C.R Doering and J.~D. Gibbons.
\newblock {\em applied analysis of the Navier-Stokes equation}.
\newblock Cambridge University Press, 1995.

\bibitem{kraichnan80}
R.H. Kraichnan and D.~Montgomery.
\newblock Two-dimensional turbulence.
\newblock {\em Reports in Progress in Physics}, 43:547--619, 1980.

\bibitem{buhler01}
U.~M\"uller L.~B\"uhler.
\newblock {\em magnetofluiddynamcis in channels and containers}.
\newblock Springer-Verlag, 2001.

\bibitem{moffatt67}
H.~K. Moffatt.
\newblock on the suppression of turbulence by a uniform magnetic field.
\newblock {\em J. Fluid. Mech.}, 28,3:571--592, 1967.

\bibitem{psm00}
A.~Poth\'erat, J.~Sommeria, and R.~Moreau.
\newblock An effective two-dimensionnal model for {MHD} flows with tranverse
  magnetic field.
\newblock {\em J. Fluid. Mech.}, 424:75--100, 2000.

\bibitem{roberts67}
P.H. Roberts.
\newblock {\em Introduction to Magnetohydrodynamics}.
\newblock Longmans, 1967.

\bibitem{sermange83}
M.~Sermange and R.~Temam.
\newblock Some mathematical questions related to the {MHD} equations.
\newblock {\em Comm. Pure Appl. Math.}, 36:635--664, 1983.

\bibitem{shebalin83}
J.~V. Shebalin, W.H. Matthaeus, and D.~Montgomery.
\newblock Anisotropy in {MHD} turbulence due to a mean magnetic field.
\newblock {\em Journal of Plasmas physics}, 29:525--547, 1983.

\bibitem{sm82}
Jo\"el Sommeria and Ren\'e Moreau.
\newblock Why, how and when, {MHD} turbulence becomes two-dimensionnal.
\newblock {\em J. Fluid Mech.}, 118:507--518, 1982.

\end{thebibliography}

\end{document}